\documentstyle[12pt]{article}
\renewcommand{\a}{\alpha}
\renewcommand{\b}{\beta}
\renewcommand{\d}{\delta}
\newcommand{\g}{\gamma}
\newcommand{\e}{\varepsilon}
\newcommand{\x}{\xi}

\newcommand{\m}{\mu}
\newcommand{\n}{\nu}

\renewcommand{\o}{\omega}
\newcommand{\D}{\Delta}
\newcommand{\G}{\Gamma}
\renewcommand{\O}{\Omega}
\renewcommand{\S}{\Sigma}
\newcommand{\p}{\psi}

\renewcommand{\t}{\tau}
\newcommand{\real}{{{\rm I} \kern -.19em {\rm R}}}
\newcommand{\tr}{{\rm {Tr} \,}}
\newcommand{\cb}{{\bar c}}
\newcommand{\half}{\frac 1 2}
\newcommand{\pa}{\partial}
\newcommand{\gdo}{g_{\mu\nu}}
\newcommand{\ie}{{{\em i.e.},\ }}
\newcommand{\cf}{{\em cf.\ }}
\newcommand{\vf}{{\varphi}}
\renewcommand{\=}{&=&} 
\newcommand{\qq}{&\qquad &}
\newcommand{\equ}[1]{(\ref{#1})}
\newcommand{\SS}{{\cal S}}
\newcommand{\wsx}[1]{\WW^{\rm S}_{(#1)}}
\newcommand{\dsx}[1]{{\delta^{\rm S}_{(#1)}}}
\newcommand{\dz}{\delta_0}
\renewcommand{\AA}{{\cal A}}
\newcommand{\BB}{{\cal B}}
\newcommand{\GG}{{\cal G}}
\newcommand{\FF}{{\cal F}}
\newcommand{\F}{{\Phi}}
\newcommand{\LL}{{\cal L}}
\newcommand{\WW}{{\cal W}}
\newcommand{\dxm}{dx^\mu}
\newcommand{\dxn}{dx^\nu}
\newcommand{\eq}{\begin{equation}}
\newcommand{\ee}{\end{equation}}
\newcommand{\eqn}[1]{\label{#1}\end{equation}}
\newcommand{\eea}{\end{eqnarray}}
\newcommand{\eqa}{\begin{eqnarray}}
\newcommand{\eqan}[1]{\label{#1}\end{eqnarray}}
\newcommand{\ba}{\begin{array}}
\newcommand{\ea}{\end{array}}
\newcommand{\eqac}{\begin{equation}\begin{array}{rcl}}
\newcommand{\eqacn}[1]{\end{array}\label{#1}\end{equation}}
\newcommand{\nes}{\nonumber\es}
\newcommand{\non}{\nonumber}
\newcommand{\Shat}{{\hat\Sigma}}
\newcommand{\BBS}{\BB_\Shat}
\newcommand{\dint}{\displaystyle{\int}}
\newcommand{\wsh}[1]{\hat\WW^{\rm S}_{(#1)}}
\newcommand{\fhat}{{\hat F}}
\newcommand{\intd}{\int d^{n+2} \! x \,\,}
\newcommand{\intr}{\int_{\real^{n+2}}}
\newcommand{\Cb}{\bar C}
\newcommand{\rangekg}{1 \leq k \leq n,\ 0 \leq g \leq n-k}
\renewcommand{\*}{*\!}
\newcommand{\wte}[1]{{\WW^{\rm T}_{(#1)}}}
\newcommand{\scr}{\scriptstyle}
\newcommand{\pad}[2]{{\displaystyle{\frac{\partial #1}{\partial #2}}}}
\newcommand{\fud}[2]{{\displaystyle{\frac{\delta #1}{\delta #2}}}}
\newcommand{\Lp}{\displaystyle{\biggl(}}
\newcommand{\Rp}{\displaystyle{\biggr)}}
\newcommand{\LP}{\displaystyle{\Biggl(}}
\newcommand{\RP}{\displaystyle{\Biggr)}}
\newcommand{\lp}{\left(}
\newcommand{\rp}{\right)}
\newcommand{\lc}{\left[}
\newcommand{\rc}{\right]}
\newcommand{\lac}{\left\{}
\newcommand{\rac}{\right\}}
\newcommand{\f}{\phi}
\newcommand{\na}{\nabla}
\newcommand{\CB}[2]{\bar C^{#1}_{#2}}
\newcommand{\ch}[2]{\chi^{#1}_{#2}}

\newcommand{\fah}[2]{\hat\f^{(#1)}_{#2}}
\newcommand{\lag}[2]{\Pi^{#1}_{#2}}
\newcommand{\bq}[2]{B^{#1}_{#2}}
\newcommand{\be}[2]{b^{#1}_{#2}}
\newcommand{\bec}[2]{\hat b^{#1}_{#2}}
\newcommand{\es}{\\[3mm]}
\newcommand{\dsum}[2]{\displaystyle{\sum_{#1}^{#2}}}
\newcommand{\dintr}{\displaystyle{\intr}}
\newcommand{\journal}[4]{{\em #1~}#2\,(19#3)\,#4;}
\newcommand{\hpa}{\journal {Helv. Phys. Acta}}
\newcommand{\ijmp}{\journal {Int. J. Mod. Phys.}}
\newcommand{\pr}{\journal {Phys. Rev.}}

\newcommand{\cmp}{\journal {Comm. Math. Phys.}}

\newcommand{\np}{\journal {Nucl. Phys.}}
\newcommand{\pl}{\journal {Phys. Lett.}}
\newcommand{\prep}{\journal {Phys. Reports}}

\newcommand{\annp}{\journal {Ann. Phys. (N.Y.)}}
\makeatletter
\@addtoreset{equation}{section}
\makeatother

\begin{document}
\begin{titlepage}
\title{
{\normalsize July 1992} \hfill {\normalsize MPI--Ph/92-57}\\
\hfill {\normalsize UGVA---DPT 1992/07--773}\\
\hfill\\
Renormalization and finiteness of\\ topological $BF$ theories
\thanks{Supported in part by the Swiss National Science Foundation.}}
\author{Claudio Lucchesi\\
{\small Max-Planck Institut f\"ur Physik}\\
{\small Werner Heisenberg Institut}\\
{\small F\"ohringer Ring 6, Postfach 40 12 12,}
{\small D -- 8000 Munich 40 (Germany)}\\
\and
\\ \\
Olivier Piguet \ \ and \ \ Silvio Paolo Sorella\\
{\small D\'epartement de Physique Th\'eorique}\\
{\small 24, quai Ernest Ansermet}\\
{\small CH -- 1211 Geneva 4 (Switzerland)}
}
\date{\quad}
\maketitle
\begin{abstract}
\noindent We show that the $BF$ theory in any space-time
dimension, when quantized in a certain linear covariant gauge,
possesses a vector supersymmetry. The generator
of the latter together with those of the BRS transformations
and of the translations form the basis of
a superalgebra of the Wess-Zumino type.
We give a general classification of all possible anomalies and invariant
counterterms. Their absence, which amounts to ultraviolet finiteness,
follows from purely algebraic arguments in the lower-dimensional cases.
\end{abstract}
\thispagestyle{empty}
\end{titlepage}
\section{Introduction}
Original inspiration for topological gauge field theory (TGFT) came from
mathe\-ma\-tics. For instance, the topology of four-dimensional
manifolds as studied by Donaldson~\cite{Donaldson} could be described
through an action principle by the topological Yang-Mills theory
of Witten~\cite{wcmp117}. Another celebrated example, in three
dimensions, is the description of invariants of knots in terms
of the topological Chern-Simons theory~\cite{s-btc,wcmp121}.

The topological Yang-Mills theory and the Chern-Simons theory are
prototypes of two distinct classes of TGFT's which are
sometimes classified as being ``of Witten type", or
respectively ``of Schwarz type" (see for instance~\cite{bbr-pr91}).
We shall only be concerned with the latter in the following.

Schwarz type TGFT's are characterized by the fact that
their classical gauge fixed action can be written as the sum of a
gauge-invariant term and a BRS-exact term. Besides
the Chern-Simons theory there exists another TGFT of Schwarz type,
namely the topological $BF$ theory. The latter represents a natural
generalization of the Chern-Simons theory since it can be defined
in arbitrary dimensions, whereas a Chern-Simons action exists only
in odd-dimensional space-times.
 Moreover, the Lagrangean of the $BF$ theory contains the quadratic
terms needed for defining a quantum theory, whereas a Chern-Simons action
 hows this feature only in three dimensions.

The topological $BF$ model describes the coupling of an
antisymmetric tensor field to a Yang-Mills potential
(\cf~\cite{bbr-pr91} for references), hence the name ``antisymmetric
tensor field theory" which is sometimes used in the literature.

The aim of the present paper is to present a systematic study
of the perturbative renormalization
of the $BF$ theory in arbitrary space-time dimensions.

Before starting with our program we feel it necessary to motivate
the recourse to perturbation theory and to compare its relevance to
the one of non-perturbative approaches to TGFT.

Non-perturbative investigations are sensitive to the topology of the
manifold on which the TGFT is defined. As an example, let us consider
the quantization law for the Chern-Simons coupling constant
$k$~\cite{quant}. This is a non-perturbative effect
that is due to the existence of topologically non-trivial finite gauge
transformations. In turn, the quantization law
enforces the vanishing of the corresponding $\b$-function $\b_k$,
which lead to the conjecture that the theory should be finite.

When dealing with perturbation theory, one sees only infinitesimal
gauge transformations, \ie gauge transformations which are connected to
the identity. In the example above, the quantization law  cannot
be inferred perturbatively.
The $\b$-function of the coupling constant has to be dealt with order
by order. This $\b$-function has been shown to be zero perturbatively
\cite{blasi-collina,dlps}.

Hence, perturbation theory is necessary for a rigorous discussion of
results obtained from non-perturbative considerations.
In our example, a perturbative analysis is necessary for
clarifying the issue of ultraviolet finiteness.
On a more abstract and fundamental level,
it is only in the perturbative regime that the
existence of the theory  may a priori  be guaranteed
by the theorems of renormalization theory~\cite{velo-wight}.

There is a further motivation for undertaking a perturbative study.
Although local observables do not exist in a TGFT,
such objects may live on the boundary, if any,
of the manifold on which the theory is defined~\cite{wcmp121,moore}.
Perturbative considerations are then expected
to lead to a better understanding of the structure of the algebra
of such local observables~\cite{blasi-col,emery}.

The ultraviolet finiteness was studied and established
for particular gauge fixings in three-dimensional Chern-Simons
theory~\cite{blasi-collina,dlps,lp-csdiff,bps-antighost}
as well as in  two-, three- and four-dimensional
$BF$ theory~\cite{maggiore-2d,maggiore,gms-plb2559165}.
In all these cases (except~\cite{blasi-collina}) the main ingredient
of the proof was the presence of a supersymmetric
structure~\cite{dgsplb22589367}. Furthermore, in most cases the
validity of a ghost equation~\cite{bps-antighost}
controlling the couplings of some Faddeev-Popov fields proved being an
important feature.

We treat in a similar way the $BF$ topological theory.
Our paper is organized as follows.
In sect. 2 we recall the main features of the topological $BF$
theory~\cite{bbr-pr91,wallet,aaf}. We choose a linear gauge fixing
condition, in contrast with~\cite{wallet} where a non-linear
covariant gauge is used. We write the corresponding
Slavnov identity, which expresses the (off-shell
nilpotent) BRS invariance. The supersymmetric structure of
the $BF$ models is worked out in sect. 3. We show that beyond the BRS
transformations there exists a vector valued
generator whose anticommutation with BRS yields the
translation generator. For a specific value of the gauge fixing
parameters this supersymmetry turns out to be an invariance of the
theory up to a trivial linear breaking. The ghost equation is derived in
sect. 4. The functional operators generating all the symmetries obey a
closed algebra displayed in sect. 5.
The problem of the renormalization is treated in sect. 6 by
cohomological methods. We determine all anomaly candidates
and all possible invariant counterterms. By anomaly candidates we mean
all the possible obstructions to the renormalization program, whereas
``all possible invariant counterterms" stands for the
ambiguities arising in the renormalization procedure,
usually also referred to as ``ultraviolet infinities".
The cases of space-time dimensions four to seven are discussed in more
details in sect. 7.

Let us remark that, the theory being topological,
its physical content should be independent
of the metric of the manifold in which it is defined. For
simplicity we shall however restrict ourselves to the case of flat
$\real^D$ space-times.
An extension of our results to arbitrary manifolds -- at least
to manifolds such that a renormalized perturbation theory can be
defined -- may be performed by following the renormalization theory
arguments of Ref.~\cite{lp-csdiff} (see also~\cite{abud-fiore} for
a formal discussion based on the path integral).
\section{BRS invariance and gauge fixing}\label{section2}
The BF system in $(d=n+2)$-dimensional space-time is
defined at the classical level by the gauge invariant action
\eq
\S_{\rm inv} = {1\over 2n!} \intd\tr \e^{\m_1\cdots\m_{n+2}}
       B_{\m_1\cdots\m_n} F_{\m_{n+1}\m_{n+2}}\ ,
\eqn{1}
where $B_{\m_1\cdots\m_n}$ is an antisymmetric tensor
of rank $n$ and $F_{\m\n}= \pa_\m A_\n - \pa_\n A_\m + [A_\m,A_\n]$
is the Yang-Mills strength. All the fields $\vf$ live in
the adjoint representation of the gauge
group $G$, which we assume to be a simple Lie group.
We use the conventions $[T_a,T_b] = i f_{abc}T_c$,
$\tr T_a T_b = \d_{ab}$, the $T_a$'s being a basis of
Lie $G$ in the fundamental representation. We also adopt the matrix
notation $\vf = \vf^a T_a$ for any field $\vf$.
In terms of the (Lie algebra valued) differential forms
$B = (1/n!)\ B_{\m_1\cdots\m_n} dx^{\m_1}\cdots dx^{\m_n}$ and
$A=A_\m \dxm$, the invariant action reads
\eq
\S_{\rm inv} = \intr\tr B \,\wedge\, F\ ,
\eqn{invaction}
with\footnote{$[X,Y]$ denotes the {\em graded commutator}, which is an
anticommutator if both $X$ and $Y$ are fermionic, and a commutator
otherwise. By fermionic, resp. bosonic fields, we mean fields with an
odd, resp. even grading, the latter being given by the sum of the
ghost number and of the form degree.}
$F=dA+1/2\ [A,A]$. We shall omit the wedge symbol
in the sequel. The field equations for $A$, resp. $B$, are
\eq
F=0\ ,\qquad\qquad DB=0\ ,
\eqn{eqmotion}
The action $\S_{\rm inv}$ is invariant under two sets
of gauge transformations, $\d_\o$ and $\d_\p$:
\eq\ba{lllllll}
\d_{\o} A \=  D\o\ , \qq \d_{\o} B \= [B,\o]\ ,\label{gaugeo}\\[3mm]
\d_{\psi} A \= 0\ ,\qq \d_\psi B \= D \psi\ ,\label{gaugep}
\ea\end{equation}
with the covariant exterior derivative $D$ given for any Lie algebra
valued form $\vf$ by $D\vf = d\vf + [A,\vf]$. Moreover
$\o=\o^aT_a$ and $\p=(1/(n-1)!)\p^a_{\m_1\cdots\m_{n-1}}T_a
dx^{\m_1}\cdots dx^{\m_{n-1}}$ are Lie G valued
forms of degree $0$, resp. $n-1$.

The Yang-Mills symmetry $\d_\o$ can be gauge fixed in the usual way
by using the Faddeev-Popov procedure. This is not the case for the
symmetry $\d_\p$ since it contains zero-modes. Indeed,
if $\p=D\p^\prime$, where $\p^\prime$ is an abritrary $(n-2)$-form,
$\d_\p B=DD\p'=[F,\p']$ which vanishes on-shell due to the equation
of motion $F=0$. $\p'=D\p''$, with $\p''$ an arbitrary $(n-3)$-form, is
then an on-shell zero-mode of the former. This procedure stops when
$\p^{(k-1)}=D\p^{(k)}$ and $\p^{(k)}$ is an arbitrary $0$-form, \ie
there are $k=n-1=d-3$ such stages. The symmetry $\d_\psi$ is said to be
$k$-th stage on-shell reducible.

Fixing a gauge symmetry is more involved when the symmetry is
reducible, as is $\d_\psi$. One possible way to do this is to
follow the Batalin-Vilkovisky (BV) prescription~\cite{bv}. Nevertheless,
the recourse to this prescription can be avoided if one applies
carefully the usual BRS renormalization prescription, \ie if one
properly fixes all the gauge freedom corresponding to the zero-modes
of the symmetry.

Hence, we shall perform here the perturbative quantization of the
$BF$ theory along the familiar lines of the BRS
renormalization procedure. We will point out the parallelism
to the BV approach.
\subsection{BRS invariance}
The first step in the BRS approach is to introduce the Faddeev-Popov
(FP) ghosts and ghosts-for-ghosts. We shall use the
notation~\cite{wallet} $B=B^0_n$, where the upper index
denotes the ghost number (or FP charge) and the lower one is the form
degree. The FP ghost $\p$ of the symmetry $\d_\p$ will be
named $B^1_{n-1}$ and its chain of ghosts-for-ghosts is
$\p'=B^2_{n-2},\ \p''=B^3_{n-3},\cdots $ up to
$\p^{(k)}=B^n_0$. The FP ghost for the Yang-Mills symmetry $\d_\o$ is
denoted by $c$. The gauge fields and the system of ghosts form the
so-called ``geometrical" sector of the theory.

In the geometrical sector, the BRS transformations write
\eq\ba{lcl}
s\, A \= Dc\ ,\\[3mm]
s\, c\= c^2\ ,\\
\ea\label{brsym}\end{equation}
and
\eq\ba{lcl}
s\, B^g_{n-g} \= DB^{g+1}_{n-g-1}+[c,B^g_{n-g}]
             \quad {\scriptstyle 0\leq g\leq n-1}\ ,\\[3mm]
s\, B^n_0 \= [c,B^n_0] \quad {\scr (g=n)}\ .
\ea\label{brsred}\end{equation}
This BRS transformation is nilpotent only {\em on-shell}, since:
\eq
s\,^2\, B^g_{n-g}=-DDB^{g+2}_{n-g-2} =-[F,B^{g+2}_{n-g-2}],
\ {\scriptstyle 0\leq g\leq n-2}.
\eqn{scarre}

Besides the geometrical fields, one has to introduce FP antighosts
and Lagrange multipliers (or Stuckelberg fields) in
order to implement the gauge fixing. For the Yang-Mills symmetry
they are denoted by $\cb$, respectively $\pi$. For the
reducible symmetry $\d_\p$ one needs the sets
of antighosts $\Cb^{\g(k)}_{n-g-k}$ and Lagrange multipliers
$\Pi^{\g(k)+1}_{n-g-k}$, for ${\scr 1\leq k\leq n\ ,\ 0\leq g\leq n-k}$
and with $\g(k)=g$ for $k$ even and $\g(k)=-g-1$ for $k$
odd\footnote{As usual the upper index denotes the ghost
number and the lower one is the form degree. We use the
notation of~\cite{wallet}.}.

The BRS transformations act on the antighosts and
on the multiplier fields as
\eq\ba{rclcrcl}
s\, \cb \= \pi\ , \qq s\, \pi\= 0\ ,\\[3mm]
s\, \Cb^{\g(k)}_{n-g-k} \= \Pi^{\g(k)+1}_{n-g-k}\ ,
\qq s\, \Pi^{\g(k)+1}_{n-g-k}\= 0\ ,\quad {\scriptstyle \rangekg}\ .
\ea\label{brscbpi}\end{equation}

The ghosts and antighosts of the reducible symmetry
form a pyramid which starts from the gauge field $B^0_n$ and
ends when its base is made out of $0$-forms:
$$
\ba{lllllllll}
&&&&B^0_n&&&&\\[3mm]
&&&\Cb^{-1}_{n-1}&&B^1_{n-1}&&&\\[3mm]
&&\Cb^0_{n-2}&&\Cb^{-2}_{n-2}&&B^2_{n-2}&&\\[3mm]
&\Cb^{-1}_{n-3}&&\Cb^1_{n-3}&&\Cb^{-3}_{n-3}&&B^3_{n-3}&\\[3mm]
...&&...&&...&&...&&...
\ea
$$
The $/$-diagonals from left to right correspond to $g=0,\ g=1,\cdots$
and the $\backslash$-diagonals from right to
left have $k=0,\ k=1,\cdots$. The diagonal with
$k=0$ contains the gauge field $B$ and its tower of ghosts.
The multiplier fields form a smaller pyramid which
corresponds to the one of the antighosts:
$$
\ba{lllllll}
&&&\Pi^{0}_{n-1}&&&\\[3mm]
&&\Pi^1_{n-2}&&\Pi^{-1}_{n-2}&&\\[3mm]
&\Pi^{0}_{n-3}&&\Pi^2_{n-3}&&\Pi^{-2}_{n-3}&\\[3mm]
...&&...&&...&&...
\ea
$$
Similarly, for the Yang-Mills symmetry, we have the (trivial) pyramids:
$$
\ba{rclcrcl}
&A& &\qquad\qquad\qquad& && \\[3mm]
\cb\qq c \qq &\pi&
\ea
$$
\subsection{Slavnov identity}\label{Id.Slavnov}
The purpose of this subsection is to extend the on-shell nilpotent
BRS operator $s$ to an operator which is nilpotent {\em off-shell}.
This is possible~\cite{carlo} by introducing into the action
terms which are non-linear in the external sources which we shall need
in order to define the renormalized BRS transformations. The off-shell
nilpotent operator will be expressed functionally by a Slavnov
identity, which will be taken as the starting
point for the definition of the theory\footnote{This
approach is similar to the one of Batalin and Vilkovisky~\cite{bv},
our Slavnov identity playing the role of their ``master
equation''. There is however an important difference: each BV
``antifield'' keeps in our approach a classical part which
we identify with the external source coupled to the
BRS variation of the corresponding field.}.

Let us thus introduce external sources
$\g=\g^{-1}_{n+1}$, $\tau=\tau^{-2}_{n+2}$ and
$\be{-g-1}{g+2},\ {\scr 0\leq g\leq n}$,
which we couple to the composite BRS variations of $A$,
$c$ and $B^g_{n-g}$, respectively.
The BRS invariance of the total classical action $\S$ is expressed
through the non-linear Slavnov identity
\eqa
\SS(\S)\equiv\intr\tr \!\!\! &&\Biggl\{
\fud{\S}{\g}\,\fud{\S}{A} + \fud{\S}{\t}\,\fud{\S}{c} +
\sum_{g=0}^n \fud{\S}{b^{-g-1}_{g+2}}\,\fud{\S}{B^g_{n-g}}
+ \,\pi\,\fud{\S}{\cb} \nonumber\\[3mm]
&& + \sum_{k=1}^{n}\sum_{g=0}^{n-k}
\Pi^{\g(k)+1}_{n-g-k}\,\fud{\S}{\Cb^{\g(k)}_{n-g-k}} \Biggr\} = 0 \ .
\label{slavnov}
\eea
The associated linearized Slavnov operator writes
\eqa
\!\!\!\!\!\SS_\S \equiv \intr\tr\!\!\! &&\!\!\! \Biggl\{
  \,\sum_{g=0}^n\, \lp \fud{\S}{b^{-g-1}_{g+2}}\,\fud{}{B^g_{n-g}}
  +\fud{ \S }{B^g_{n-g}}\,\fud{}{b^{-g-1}_{g+2}} \rp
  +\fud{ \S }{\g}\,\fud{}{A} + \fud{\S }{A}\,\fud{}{\g}
  \nes
&&\!\!\!+ \fud{\S}{\t}\,\fud{}{c} + \fud{\S}{c}\,\fud{}{\t}
     + \,\pi\,\fud{}{\cb} +  \dsum{k=1}{n} \dsum{g=0}{n-k}
\Pi^{\g(k)+1}_{n-g-k}\,\fud{}{\Cb^{\g(k)}_{n-g-k}} \Biggr\} = 0\ .
\label{slavnov-lin}\eea
$\SS_\S$ is the {\em off-shell} nilpotent extension of the BRS
operator $s$ we were looking for. Indeed,
$\SS_\S^{\,\,\, 2}=0$ if the Slavnov identity \equ{slavnov} is
satisfied.
\subsection{Gauge fixing}
Within the functional formalism, we fix the gauge by
imposing the following gauge conditions:
\eqa
\fud{\S}{\pi} \= d\* A\ ,\nes
\fud{\S}{\lag{-g}{n-g-1}} \= (-1)^{n+1} d\* \bq{g}{n-g}
   + (-1)^{n+1+g} \* d\CB{g}{n-g-2}\nes
&& +\, x_{1,g,n} \*\lag{g}{n-g-1}\ ,
   \quad{\scr 0\leq g\leq n-1,\ (k=1)}\ , \label{gauge-cond}\es
\fud{\S}{\lag{\g(k)+1}{n-g-k}}\= (-1)^{n+1} d\*\CB{\g(k-1)}{n-g-(k-1)}
   + (-1)^{n+k+g} \* d\CB{\g(k+1)}{n-g-(k+1)}\nes
&& + \lac \ba{ll} (-1)^{g+1}x_{k-1,g+1,n} & {\scr (k\ {\rm even})}\\
                  x_{k,g,n}       & {\scr (k\ {\rm odd})}
          \ea
     \rac
   \*\lag{\g(k-1)}{n-g-k}\ ,
   \quad {\scr  2\leq k\leq n,   \ 0\leq g\leq n-k}\ , \non
\eea
where the gauge parameters $x_{k,g,n}$ are (for the
moment free) numerical coef\-fici\-ents\footnote{Integrability
of the equations has been enforced.}$\phantom{}^{,}$\footnote{The symbol
$*$ denotes the Hodge duality, defined for any $p$-form $\o$ by
\[
*\o = {1\over (d-p)!}\ (*\o)_{\b_1\b_2\cdots\b_{d-p}}
      \ dx^{\b_1}dx^{\b_2}\cdots dx^{\b_{d-p}}\ ,
\]
where
\[
(\ast\o)_{\b_1\b_2\cdots\b_{d-p}} = {1\over p!}\,\,\e_{\b_1\b_2\cdots
\b_{d-p}\a_1\cdots\a_p}\ \o^{\a_1\cdots\a_p}\ .
\]
}.

We want to find the most general classical action $\S$ obeying the
Slavnov identity \equ{slavnov} and the gauge conditions
\equ{gauge-cond}. One easily sees that  the compatibility
of these  two requirements implies the antighost equations
\eqa
\lp\fud{}{\cb}-d\*\fud{}{\g}\rp\S \= 0\ ,\nes
\lp\fud{}{\CB{-g-1}{n-g-1}}-d\*\fud{}{\be{-g-1}{g+2}}\rp\S
  \= (-1)^g\* d\lag{g+1}{n-g-2}\ ,
  \qquad {\scr 0\leq g\leq n-1,\ (k=1)}\ ,\label{antighosteq}\es
\fud{}{\CB{\g(k)}{n-g-k}}\,\S \= d\*\lag{\g(k-1)+1}{n-g-(k-1)}
  + (-1)^{k+g+1} \* d\lag{\g(k+1)+1}{n-g-(k+1)} \ ,\nes
&&{\scr 0\leq g\leq n-k,\ 2\leq k\leq n}\ .\non
\eea

Let us first write down the general solution of
the gauge conditions \equ{gauge-cond} and of the
antighost equations \equ{antighosteq}; we get
\eq
\S(A,c,\bq{}{},\cb,\pi,\CB{}{},\lag{}{},\g,\tau,\be{}{})
=\Shat(A,c,\bq{}{},\hat\g,\tau,\bec{}{})
 +\S_{\Pi}(A,B,\cb,\pi,\Cb,\Pi)\ ,
\eqn{shatspi}
where
\eqa
\S_{\Pi} =&& \intr\tr\Bigl\{ -d\pi \* A
 -\dsum{g=0}{n-1} d\Pi^{-g}_{n-g-1}\* B^g_{n-g}     \nes
&&+\ \dsum{k=2}{n}\,\dsum{g=0}{n-k}
 \Bigl[-d\Pi^{\g(k)+1}_{n-g-k}\*\Cb^{\g(k-1)}_{n-g-k+1}
 +(-1)^{n+1}\,d\Cb^{\g(k)}_{n-g-k}\*\Pi^{\g(k-1)+1}_{n-g-k+1}
 \Bigr] \nes
&&+\ \dsum{k=1,3,5,\cdots}{n}\ \dsum{g=1}{n-k}\,x_{k,g,n}\,
   \Pi^{-g}_{n-g-k}\*\Pi^g_{n-g-k}\Bigr\}\ . \label{sigmapi}
\eea
The truncated action $\Shat$ is the general solution of the homogenous
gauge conditions and of the homogenous antighost equations. It is
hence independent of the multiplier fields and depends on the
antighosts only through the combinations
\eqa
\hat\g \= \g -  \*\, d\cb\ ,\nes
\bec{-g-1}{g+2} \= \be{-g-1}{g+2} + (-1)^{g+1} \* d \CB{-g-1}{n-g-1}\ ,
           \quad {\scr 0\leq g\leq n-1,\ (k=1)}\ ,\label{bhat}\es
\bec{-n-1}{n+2} \= \be{-n-1}{n+2}
           \quad {\scr (g=n,\ k=1)}\ .\non
\eea

The part of the action which depends on the multipliers
completely fixes the gauge freedom.
Apart from the terms quadratic in the multiplier fields
it is a gauge-non-covariant version of the gauge fixing of
Ref.~\cite{wallet}. The coefficients $x_{k,g,n}$ of the
quadratic terms are still free and will remain so after
imposing the Slavnov identity. It is only the requirement of
supersymmetry that will enforce a definite value for these
coefficients, see sect. \ref{susy}.

Before writing the Slavnov identity for the truncated action $\Shat$,
let us remark that there exists a very compact notation\footnote{An
equivalent notation is found in Ref.~\cite{aaf}}
for the fields appearing as arguments of $\Shat$.
These can be arranged within two ``field ladders''
\eqac
\fah{1}{} \= \{\fah{1}{p},\ {\scr p=0,\cdots,n+2} \}
           = \{ c,\ A,\ \bec{-g-1}{g+2}\ {\scr (g=0,\cdots,n)} \}\ ,\es
\fah{2}{} \= \{\fah{2}{p},\ {\scr p=0,\cdots,n+2} \}
           = \{\bq{g}{n-g}\ {\scr (g=n,\cdots,0)},\ \hat\g, \t \} \ .
\eqacn{ladder-hat}
Each of these ladders contains forms of degrees  ranging from $0$ up to
the maximal degree $n+2$; we shall call such ladders ``complete".
In these ladder variables, and after use of the gauge
conditions and of the antighost equations, the Slavnov identity
\equ{slavnov} takes the form\footnote{One recognizes here the
``master equation'' of Batalin and Vilkovisky~\cite{bv}. The
fields \equ{bhat} play the role of the BV ``antifields'', but they
possess here a classical part which has no analogue in the BV
formalism and which consists of our external sources $\g$, $b$
and $\t$ (for $\t$ the shift \equ{bhat} is trivial: $\hat\t =\t$).}
\eq
\SS(\S)=\half\BBS\Shat=0\ ,
\eqn{slavnov-hat}
where we define, for any functional $\g$, the $\g$-dependent
functional linear Slavnov operator $\BB_\g$ to be
\eq
\BB_{\g} \equiv \intr\tr\sum_{p=0}^{n+2}\lp
    \fud{\g}{\fah{2}{n+2-p}}\fud{}{\fah{1}{p}}
  + \fud{\g}{\fah{1}{n+2-p}}\fud{}{\fah{2}{p}} \rp\ .
\eqn{bbs}

The general solution of the Slavnov identity \equ{slavnov-hat} reads, up
to trivial field renormalizations,
\eq
\Shat = \dintr\tr \lp \dsum{p=0}{n+1} \fah{1}{p}d\fah{2}{n+1-p}
   + \half\dsum{p=0}{n+2}\,\dsum{q=0}{\,n+2-p}
        \lc\fah{1}{p},\fah{1}{q}\rc\fah{2}{n+2-p-q} \rp
\eqn{hat-action}
The cubic part contains terms quadratic in the external sources,
and hence terms quadratic in the antighosts. The need for such
non-linear terms was stressed at the beginning of
subsect.~\ref{Id.Slavnov}.

In the case where the functional $\g$ in $\BB_\g$ \equ{bbs}
is the truncated action  $\Shat$ given in \equ{hat-action},
then $\BB_\g$ is nilpotent {\em off-shell}, and we shall write
\eq
\BBS \equiv \BB \ , \qquad \BB^2=0 \ .
\eqn{bb-class}
The {\em off-shell nilpotent} BRS transformations
of the  components of the ladders \equ{ladder-hat} are
now given by
\eq\ba{l}
\BB\fah{1}{p} = d\fah{1}{p-1}
      + \half \dsum{q=0}{n+2}\lc\fah{1}{q},\fah{1}{p-q}\rc \ ,
      \quad {\scr p=0,...,n+2}\ ,\es
\BB\fah{2}{p} = d\fah{2}{p-1}
      + \phantom{\half}\dsum{q=0}{n+2}\lc\fah{1}{q},\fah{2}{p-q}\rc \ ,
      \quad {\scr p=0,...,n+2}\ .
\ea\eqn{brs-ladder}
\section{Supersymmetry}\label{susy}
In analogy to what is known for the case of the Chern-Simons theory
in three dimensions~\cite{dgsplb22589367,dlps,bbr-pr91}, the topological
$BF$ theory exhibits a supersymmetric structure which, for the time
being, has only been established in dimensions four and less
\cite{gms-plb2559165,maggiore,maggiore-2d}.
In the present section, we shall give a systematic
treatment of this supersymmetry in higher dimensions,
at the classical level. The problems faced by its
renormalization are addressed in sect. \ref{renormalisation}.

Let the following transformations of the ladder components
\equ{ladder-hat} be
\eq\ba{rcl}
\dsx{\x} \fah{A}{p} \=
\left\{\ba{ll}-i_\x\,\fah{A}{p+1} \ , \quad&{\scr p=0,...,n+1} \ ,\es
              0 \ , \quad&{\scr p=n+2}\ ,
       \ea
\right.\qquad A=1,2\ ,
\ea\eqn{susy-ladder}
where $i_\x$ is the inner derivative along a (constant)
vector field $\x^\m$. In the sector of the antighosts and multipliers,
for the Yang-Mills symmetry, let
\eq
\dsx{\x}\,\cb=0\ ,\qquad\dsx{\x}\pi=\LL_\x\cb\ ,
\eqn{susy-ym}
and, for the reducible symmetry, with ${\scr \rangekg}$, let
\eqac
\dsx{\x}\,\Cb^{\g(k)}_{n-g-k}\=\left\{
   \ba{ll}
   (-1)^{n+(k+1)/2}\,g(\x)\,\Cb^{-g-2}_{n-g-k-1}\ ,
                                        &{\scr (k\ {\rm odd})},\\[3mm]
   (-1)^{k/2+1}\,i_\x\,\Cb^{g-1}_{n-g-k+1}\ ,&{\scr (k\ {\rm even})},
   \ea\right.\\[8mm]
\dsx{\x}\,\Pi^{\g(k)+1}_{n-g-k}\=\left\{
   \ba{ll}
   (-1)^{n+(k+1)/2}\,g(\x)\,\Pi^{-g-1}_{n-g-k-1}
   +\LL_\x\,\Cb^{-g-1}_{n-g-k}\ ,
    &{\scr (k\ {\rm odd})}\ ,\\[3mm]
   (-1)^{k/2+1}\,i_\x\,\Pi^g_{n-g-k+1}
   +\LL_\x\,\Cb^g_{n-g-k}\ ,                 &{\scr (k\ {\rm even})}\ .
   \ea\right.
\eqacn{susy-red}
$\LL_\x=[i_\x,d]$ is the Lie derivative along $\x^\m$.
$g(\x)=\gdo\,\x^\m\,\dxn$ is a $1$-form of ghost number $1$,
the vector field $\x^\m$ carriying an odd grading. $\gdo$
is a flat, Minkowskian or Euclidean metric in $(n+2)$-spacetime.

At the functional level, the invariance under the transformations
$\dsx{\x}$ given by eqs. \equ{susy-ladder} to \equ{susy-red}
is implemented by the Ward operator
\eq
\wsx{\x} \equiv \intr \tr \sum_{{\rm all\ fields}\ \varphi}
\dsx{\x}\varphi\,\fud{}{\varphi}\ ,
\eqn{wardops}

The transformations generated by the operator $\wsx{\x}$ form,
together with the off-shell nilpotent linear Slavnov operator
\equ{slavnov-lin} an algebra which closes
{\it off-shell} on the space-time translations:
\eq
[\SS_\S,\wsx{\x}]=\wte{\x}\ ,
\qquad {\SS_\S\,}^2=0\ ,\qquad [\wsx{\x},\wsx{\x'}]=0\ ;
\eqn{algsd}
we denote by $\wte{\e}$ the Ward operator for translations
along a (constant) vector field $\e^\m$:
\eq
\wte{\e} \equiv \intr\tr\sum_{\rm all\ fields\ \vf}
\LL_\e\vf\,\fud{}{\vf}\ .
\eqn{wtee}

Due to the algebraic structure \equ{algsd}, we call
``supersymmetry transformations" the rules \equ{susy-ladder}
to \equ{susy-red}. The bosonic degrees of freedom are just the field
components with even grading and the fermionic ones are those with odd
grading. There is, in all dimensions, an equal number of
bosonic components and of fermionic ones\footnote{Note that
the grading of some of the fields depends on the space-time
dimension.}.

The Ward identity of this supersymmetry contains at the classical
level a  breaking which, being linear in the quantum fields,
does not get renormalized. This is analogous to what has been
proven for the Chern-Simons theory \cite{dlps}. We get
\eq
\wsx{\x}\,\S = \D^{\rm S}_{(\x)\ {\rm class}}\ ,
\eqn{wards}
where the classical breaking $\D^{\rm S}_{(\x)\ {\rm class}}$
is given by
\eqa
\D^{\rm S}_{(\x)\ {\rm class}}
=\intr\tr\Biggl\{\!\!&&\!\! -\sum_{g=0}^n b^{-g-1}_{g+2}
\,\LL_\x B^g_{n-g}+(-1)^n\,\g\,\LL_\x A
+(-1)^n\,\t\,\LL_\x c \nes
&&\!\! + d\* i_\x \,(\,b^{-1}_2 \pi + \g\Pi^0_{n-1}\,)\,\Biggr\}\ ,
\label{classbr}
\eea
provided the classical action $\S$ is just the expression given by
\equ{shatspi}, \equ{sigmapi} and \equ{hat-action}, with
fixed coefficients
\eq
x_{k,g,n}=(-1)^{n+g+(k+1)/2}\ .
\eqn{xkgn}

On the truncated action $\Shat$ \equ{hat-action},
the Ward identity of supersymmetry becomes
\eq
\hat\wsx{\x}\,\Shat = \hat\D^{\rm S}_{(\x)\ {\rm class}}\ ,
\eqn{wardsusyshat}
where the Ward operator writes, in the ladder notation \equ{ladder-hat},
\eq
\hat\wsx{\x} \equiv \intr \tr \sum_{A=1,2}\,\sum_{p=0,...,n+1}\,
-i_\x\,\fah{A}{p+1}\,\fud{}{\fah{A}{p}}\ ,
\eqn{wardopsusyhat}
and the classical breaking is replaced by
\eq
\hat\D^{\rm S}_{(\x)\ {\rm class}}=\intr\tr\lp -\sum_{p=0}^{n+2}
\fah{2}{n+2-p}\,\LL_\x\,\fah{1}{p} \rp\ .
\eqn{classbrladderhat}
\section{The ghost equation}\label{equationdughost}
It is known that, in gauge theories with a Landau gauge fixing, the
dependence of the radiative corrections on the Faddeev-Popov ghosts is
constrained by identities called the ghost
equations~\cite{bps-antighost}. Since in the present case the gauge
fixing of the field $A$ is precisely of the Landau type, one may expect
such an identity to hold. Inspection of the classical
action (given in \equ{shatspi},\equ{sigmapi}
and \equ{hat-action}) shows that this is indeed the case for
the ghost zero-form $\bq{n}{0}$. This ghost
equation reads, at the classical level:
\eq
\GG\S \equiv \intr \lp \fud{\S}{\bq{n}{0}}
   + \lc \fud{\S}{\pi},\;\CB{-n}{0}\rc  \rp = \D_{\rm class}^\GG \ ,
\eqn{ghosteq}
where the right-hand-side
\eq\ba{rl}
\D_{\rm class}^\GG \equiv \dintr \LP&  \frac{1}{2}\dsum{p=2}{n}
                            \lc \be{-p+1}{p},\;\be{-n+p-1}{n+2-p}\rc
         + \lc \be{-n}{n+1},\;A\rc + \lc \be{-n-1}{n+2},\;c\rc  \es
     &+\dsum{p=2}{n}(-1)^{n+p+1}
              \lc \be{-p+1}{p} ,\; \*\,d\CB{-n+p-1}{p-1}\rc \RP\ ,
\ea\eqn{deltag}
is linear in the quantum fields, hence not subject to renormalization.

One can check that, for any functional $\g$,
\eq
\GG\SS (\g) + \SS_\g (\GG\g-\D_{\rm class}^\GG) =
\FF\g-\D_{\rm class}^\FF\ ,
\eqn{ghost,slavnov}
with
\eq\ba{rl}
\FF \equiv \dintr & \LP
   \lc c,\;\fud{}{\bq{n}{0}}\rc +\lc A,\;\fud{}{\bq{n-1}{1}}\rc
    + \dsum{p=2}{n} \lc \be{-p+1}{p},\;\fud{}{\bq{n-p}{p}}\rc \es
  & + \lc \be{-n}{n+1},\;\fud{}{\g}\rc
    + \lc \be{-n-1}{n+2},\;\fud{}{\tau}\rc
    + \dsum{p=2}{n} \lc \*\, d\CB{-p+1}{n-p+1},\;\fud{}{\bq{n-p}{p}}\rc
                                                               \es
  & + (-1)^{n+1}  \lc \CB{-n}{0},\;\fud{}{\cb}\rc
    + (-1)^{n+1}  \lc \lag{-n+1}{0},\;\fud{}{\pi}\rc \RP \ ,
\ea \eqn{op-f}
and
\eq
\D_{\rm class}^\FF \equiv \dsum{q=2}{n} (-1)^{n+p+1}
              \lc\be{-p+1}{p},\;\*\, d\lag{-n+p}{p-1}\rc  \ .
\eqn{deltaf}

For $\g=\S$, solution of the Slavnov   identity and of the ghost
equation, \equ{ghost,slavnov} implies the ``associated ghost equation''
\eq
\FF\S = \D_{\rm class}^\FF \ ,
\eqn{ghost-assoc}
whith again a right-hand-side which is linear in the quantum fields.

In term of the truncated action \equ{hat-action} and the ladder
variables \equ{ladder-hat} the ghost equation and
its associated equation take the much simpler forms
\eq
\hat\GG\Shat \equiv \dintr \fud{\Shat}{\bq{n}{0}} =
   \half\intr\dsum{p=0}{n+2} \lc\fah{1}{p},\;\fah{1}{n+2-p}\rc
    \equiv \hat\D_{\rm class}^\GG          \ ,
\eqn{ghosteq-hat}

\eq
\hat\FF\Shat \equiv
 \dintr\dsum{p=0}{n+2} \lc\fah{1}{p},\;\fud{\Shat}{\fah{2}{p}}\rc = 0 \ .
\eqn{ghost-assoc-hat}
\section{Functional algebra}\label{algebre}
The Slavnov operator $\SS$ \equ{slavnov} and its linearized form
$\SS_F$ (see \equ{slavnov-lin}), the Ward identity
operators\footnote{The infinitesimal supersymmetry
parameters $\x^\m$ are anti\-com\-muting num\-bers, the translation
parameters $\e^\m$ are commuting.}
for supersymmetry $\wsx{\x}$ \equ{wardops} and for translations
$\wte{\e}$ \equ{wtee}, as well as the ghost equation operator
$\GG$ \equ{ghosteq} and its associated operator $\FF$ \equ{op-f}
obey the non-linear algebra
\eq\ba{l}
\SS_F\SS(F) = 0\ ,\es
\wsx{\x}\SS(F) - \SS_F \lp \wsx{\x}F-\D^{\rm S}_{(\x)\,{\rm class}} \rp
          = \wte{\x}F                         \ ,\es
\wte{\e}\SS(F) - \SS_F\wte{\e}F = 0               \ ,\es
\GG\SS(F) -(-1)^n\SS_F\lp \GG F-\D^\GG_{\rm class}\rp
         = \FF F-\D^\FF_{\rm class}      \ , \es
\FF\SS(F) +(-1)^n\SS_F\lp \FF F-\D^\FF_{\rm class}\rp  = 0\ ,
\ea\eqn{alg-nonlin}
valid for any functional $F$, and the (anti)-commutation
relations\footnote{Trivial commutation relations involving the
translation operator are not written.}

\eq\ba{l}
\lc\wsx{\x},\wsx{\x'}\rc= 0\ ,\es
\lc\GG,\wsx{\x}\rc = \lc\FF,\wsx{\x}\rc  = 0\ , \es
\lc\GG^a(x),\GG^b(y)\rc
   = \lc\FF^a(x),\GG^b(y)\rc =\lc\FF^a(x),\FF^b(y)\rc  = 0 \ .
\ea\eqn{alg-lin1}

For a functional $\hat F$ independent of the Lagrange multiplier fields
and depending on the antighosts only through the shifted external fields
\equ{bhat} -- like the truncated action $\Shat$ defined in
\equ{shatspi} -- the non-linear algebra \equ{alg-nonlin} becomes
\eq\ba{l}
\BB_\fhat\BB_\fhat\fhat = 0\ ,\es
\half\wsh{\x}\BB_\fhat\fhat
    - \BB_\fhat \lp \wsh{\x}\fhat-\hat \D^{\rm S}_{(\x)\,{\rm class}} \rp
          = \wte{\x}\fhat                         \ ,\es
\half\wte{\e}\BB_\fhat\fhat - \BB_\fhat\wte{\e}\fhat = 0           \ ,\es
\half\hat\GG\BB_\fhat\fhat
    -(-1)^n\BB_\fhat\lp \hat\GG\hat F-\hat\D^\GG_{\rm class}\rp
         = \hat\FF\hat F-\hat\D^\FF_{\rm class}      \ , \es
\half\hat\FF\BB_\fhat\fhat  +(-1)^n\BB_\fhat \hat\FF\hat F  = 0\ .
\ea\eqn{alg-nonlin2}
$\BB_\g$ was defined in \equ{bbs}, $\wsh{\x}$, $\hat\D^{\rm S}_{(\x)\,
{\rm class}}$ in \equ{wardopsusyhat}, \equ{classbrladderhat},
and $\hat\GG$, $\hat\FF$, $\hat\D^\GG_{\rm class}$ in
\equ{ghosteq-hat}, \equ{ghost-assoc-hat}. The operators  $\wsh{\x}$,
$\hat\GG$ and $\hat\FF$ obey the same linear algebra \equ{alg-lin1}
as the unhatted ones.

Furthermore if this functional $\hat F$ is the classical truncated action
$\Shat$  \equ{shatspi}, \equ{hat-action}, solution of the Slavnov
identity \equ{slavnov-hat} (and more generally if $\hat F$ obeys
the latter Slavnov identity), then to \equ{alg-nonlin}
there corresponds the linear algebra
\eq\ba{ll}
\BB^2=0\ ,& \es
\lc\wsh{\x},\BB\rc = \wte{\x}  \ ,&  \lc\wte{\e},\BB\rc=0\ ,\es
\lc \hat\GG,\BB\rc = \hat\FF\ , & \lc\hat\FF,\BB\rc = 0 \ ,
\ea\eqn{alg-lin2}
with $\BB$ defined by \equ{bbs}, \equ{bb-class}.
\section{Renormalization}\label{renormalisation}
\subsection{Statement of the problem}
Our aim is now to perform the perturbative quantization of the
classical theory defined in the preceding sections. This means that
we must:
\begin{description}
\item[1)] construct a vertex functional $\G= \S+ O(\hbar)$ satisfying
to all orders of perturbation theory all the functional
identities, in particular the Ward identities,
which we have shown to hold for the classical action $\S$.
This is the problem of the anomalies;
\item[2)] look for the possible counterterms which one can freely add
at each order to the action without spoiling the functional identities.
This is the problem of the invariant counterterms.
\end{description}
Let us begin with the problem of the anomalies and let us denote by
$\D^{{\rm BRS}}$, $\D^{\rm S}_{(\x)}$, $\D^{\GG}$ and
$\D^{\FF}$ the possible radiative breakings of the Slavnov identity
\equ{slavnov}, of the supersymmetry Ward identity \equ{wards}, of the
ghost equation \equ{ghosteq} and of its  associated equation
\equ{ghost-assoc}. From the quantum action principle~\cite{action-pr}
the breakings are local insertions of dimensions constrained
by power-counting. Their ghost numbers are fixed by ghost number
conservation.

The renormalization scheme is assumed to preserve Poincar\'e invariance.
The renormalizability of the gauge fixing conditions
\equ{gauge-cond} and of the antighost equations \equ{antighosteq}
is trivial; we therefore assume these to
hold exactly. As a consequence, the vertex functional $\G$ is
the sum of a term $\hat\G$ depending only on the ladder fields
\equ{ladder-hat}, and of the explicit term $\S_\Pi$
 depending on the Lagrange multiplier and antighost
fields which appears in the classical action \equ{shatspi}. We have:
\eq
\G(A,c,\bq{}{},\cb,\pi,\CB{}{},\lag{}{},\g,\tau,\be{}{})=
  \hat\G(\fah{1}{},\fah{2}{}) +
\S_\Pi(A,\bq{}{},\cb,\pi,\CB{}{},\lag{}{})\ .
\eqn{vertex-hat}

As a consequence of the gauge conditions and of the antighost
equations the radiative breakings depend only on the ladder fields.
Due to the algebra \equ{alg-nonlin2} (with $\hat F=\hat\G$) and
\equ{alg-lin1} the breakings obey, at the lowest order $N$ in $\hbar$
at which they are supposed to appear, the consistency
conditions\footnote{Poincar\'e invariance of the renormalization scheme
being assumed, there is no radiative breaking of translation
invariance. Hence the corresponding  consistency conditions reduce
to the condition of translation invariance of the other breakings.}
\eq\ba{ll}
\BB \D^{{\rm BRS}} =0\ ,
              & \wsh{\x}\D^{{\rm BRS}} - \BB \D^{\rm S}_{(\x)} = 0\ , \es
\hat\GG\D^{{\rm BRS}} + (-1)^{n+1}\BB\D^{\GG}= \D^{\FF}\ ,
             &\hat\FF\D^{{\rm BRS}} + (-1)^{n}\BB\D^{\FF}=0\ ,\es
\wsh{\x}\D^{\rm S}_{(\x')} - \wsh{\x'}\D^{\rm S}_{(\x)} = 0\ ,
             &\hat\GG\D^{\rm S}_{(\x)} - \wsh{\x}\D^{\GG}=0\ ,\es
\hat\FF\D^{\rm S}_{(\x)} - \wsh{\x}\D^{\FF}=0\ ,
 &\hat\GG_a\D^{\GG}_b  + (-1)^{n+1} \hat\GG_b\D^{\GG}_a=0\ ,\es
\hat\GG_a\D^{\FF}_b  - \hat\FF_b\D^{\GG}_a=0\ ,
 &\hat\FF_a\D^{\FF}_b  + (-1)^{n} \hat\FF_b\D^{\FF}_a=0\ .
\ea\eqn{consistency}
Anomalies are non-trivial solutions of these consistency conditions, \ie
solutions wich cannot be written as:
\eq\ba{llll}
\D^{{\rm BRS}} = \BB\D\ , & \D^{\rm S}_{(\x)}=\wsh{\x}\D\ ,
      &  \D^{\GG}=\hat\GG\D\ ,& \D^{\FF}=\hat\FF\D\ ,
\ea\eqn{trivial-sol}
$\D$ being some local functional of the ladder fields \equ{ladder-hat}.
In the anomaly-free situation, adding $-\D$ to the action as a
counterterm ensures the validity of the functional identities at
the order $N$ considered.

Let us now come to the invariant counterterms. Here the problem
consists in finding the general solution $\D^{\rm inv}$ -- a local
functional with the dimensions of the classical action and of ghost
number zero -- of the equations
\eq\ba{llll}
\BB\D^{\rm inv}=0\ , & \wsh{\x}\D^{\rm inv}=0\ ,
      & \hat\GG\D^{\rm inv}=0\ ,& \hat\FF\D^{\rm inv}=0\  .
\ea\eqn{invariants}
This yields all the possible local counterterms which one may recursively
add to the action without spoiling the functional identities.
\subsection{Cohomology}
We have to solve the consistency conditions \equ{consistency} and
the invariance conditions \equ{invariants} in
the space of translation invariant local functionals.
Since the translation operator $\wte{\e}$ \equ{wtee} belongs
to the algebra of functional operators of sect. \ref{algebre}
in a non-trivial way -- it appears in a right-hand side
-- it will turn out to be convenient to keep it among the set of
functional operators.

The whole set of functional operators can be incorporated
into one single operator
\eq
\d= \BB + \wsh{\x} +\wte{\e} + \tr( u\hat\GG + v\hat\FF)
        +\x^\m \pad{}{\e^\m} + \tr \lp u\pad{}{v}\rp \ ,
\eqn{cob}
where the ``global ghosts'' $\x^\m$, $\e^\m$, $u^a$ and $v^a$ are the
infinitesimal parameters of the supersymmetry transformations,
of the translations, of the ghost equation and of its
associated equation \equ{ghost-assoc}. Their  ghost numbers are 2, 1,
$n$, and $n-1$ , respectively, so that $\d$ has ghost number one. Their
gradation is equal to the parity of their ghost numbers:
for $\x$ and $\e$ it is opposite to the natural
gradation used previously. Their own transformation
laws are given by the last two terms in \equ{cob}; this
together with the algebra \equ{alg-lin1}, \equ{alg-lin2} makes $\d$ a
coboundary operator~\cite{lp-csdiff}:
\eq
\d^2=0\ .
\eqn{nilpotency}

In this cohomological setting the problems of the anomalies
and of the invariant counterterms, as described above,
reduce to one single cohomology problem. Indeed,
both the consistency conditions \equ{consistency} and the invariance
conditions \equ{invariants} can be written as
\eq
\d\D^G_{(n+2)} = 0\ ,\qquad {\scr G=0,1}\ ,
\eqn{cocyceq}
where $\D^G_{(n+2)}$ belongs to the space of integrated
local functionals of ghost number $G$ of the ladder fields
$\fah{1}{}$ and $\fah{2}{}$ \equ{ladder-hat}.
\begin{description}
\item[1)] The possible anomalies, obeying the consistency conditions
\equ{consistency}, are given by the non-trivial solutions
of \equ{cocyceq} with ghost number $G=1$ and which are homogeneous
of degree one in the infinitesimal parameters.
\item[2)] The invariant counterterms, obeying the conditions
\equ{invariants}, are obtained as the general solution of
\equ{cocyceq} with ghost number $G=0$, independent of the
infinitesimal parameters $\x^\m$, $\cdots$, $v^a$. The
non-trivial solutions correspond to the renormalization of the physical
parameters of the theory.
\end{description}
We have thus to solve the cohomology of the coboundary operator
$\d$, \ie to look for the equivalence classes modulo--$\d$
of solutions of the equation \equ{cocyceq}. ``Modulo--$\d$''
means up to a ``trivial'' term of the form $\d\hat\D^{G-1}$ with
$\hat\D^{G-1}$ taken in the same space of functionals as $\hat\D^G$.

In order to solve this cohomology problem it is useful to introduce
the filtering operator
\eq
F =
 \x^\m\pad{}{\x^\m} + \e^\m\pad{}{\e^\m} + \tr \lp u\pad{}{u}
                    + v\pad{}{v}\rp  \ .
\eqn{filter}
The insertion $\D^G_{(n+2)}$ and the operator $\d$
can be expanded according to the eigenvalues $f$ of
the filter $F$, \ie according to the degree in the global ghosts.
The expansion of $\d$ reads
\eq\ba{rl}
&\d= \dz + \d_1 \ ,\es
{\rm with:}&\quad
   \dz= \BB + \x^\m \pad{}{\e^\m} + \tr \lp u\pad{}{v}\rp\ ,\es
          &\quad
   \d_1 = \wsh{\x} + \wte{\e} + \tr( u\hat\GG + v\hat\FF)\ .
\ea\eqn{d0+d1}
One has
\eq
\dz^2=\{\dz,\d_1\}=\d_1^2=0\ .
\eqn{nilpz}
The reason for performing this filtration is that the cohomology of $\d$
is isomorphic to a subspace of the cohomology of $\dz$
(see~\cite{piguetsibold86,BBBCDcmp12088121}). Let us hence begin by
solving
\eq
\dz\D^G_{(n+2)} = 0 \ .
\eqn{cobz}

We first observe that the global ghosts $\e$, $\x=\dz\e$, $v$ and
$u=\dz v$ form two $\dz$-doublets. It
follows~\cite{piguetsibold86,brandt} that the $\dz$-cohomology
does not depend on them. Accordingly $\D^G_{(n+2)}$ will be
assumed to depend only on the field ladders \equ{ladder-hat}
$\fah{A}{}$, $A=1,2$.

Let us write
\eq
\D^G_{(n+2)} = \intr \ch{G}{n+2} \ ,
\eqn{deltaq}
where the upper index indicates  as usual the ghost number
and the lower one the form degree. The condition \equ{cobz}
implies that the $\dz$-variation of the integrand $\ch{G}{n+2}$ is
a total derivative $d\ch{G+1}{n+1}$. Then, nilpotency of $\dz$ together
with the triviality~\cite{bonora,brandt} of the cohomology
of the exterior derivative $d$ in the space of local
functionals, imply the following set of descent equations:
\eq\ba{l}
\dz\ch{g}{n+2+G-g} = d\chi^{g+1}_{n+1+G-g} \ ,
                     \qquad {\scr g = G,\cdots, G+n+1} \ ,\es
\dz\ch{n+2+G}{0} = 0\ ,
\ea\eqn{descent}
which relate $\ch{g}{n+2}$ to a zero-form of ghost number $n+2+G$.
The most general expression for this zero-form $\ch{n+2+G}{0}$  is a
polynomial in the ladder fields $\fah{A}{0}$.

Let us first consider the case of space-time dimensions greater
or equal to $5$, \ie $n\geq3$. The general solution $\ch{n+2+G}{0}$
of the last equation \equ{descent} (for $G=$ 0 or 1) is an arbitrary
linear superposition of the monomials~\cite{bonora,brandt}
\eq
{\chi'\,}^{n+2+G}_0  = \left. \prod_{k=1}^K
                        \tr c^{n_k}
          \right|_{\sum_k n_k=n+2+G\ ,\quad n_k\ {\rm odd}} \ ,
\eqn{chi0}
and of the monomial (for $G=0$)
\eq
{\chi''\,}^{n+2}_0 = \tr ( c^2\bq{n}{0})
     = -\dz \tr (c\bq{n}{0}) \ .
\eqn{chi'0}

For $n=2$~\cite{maggiore} the general solution is a superposition of:
\eq
{\chi'\,}^5_0  =  \tr c^5 \qquad  (G=1)\ ,
\eqn{chi0-2}
\eq
{\chi''\,}^4_0 = \tr (c^2\bq{2}{0})
     = -\dz \tr (c\bq{2}{0}) \qquad (G=0) \ ,
\eqn{chi'0-2}
and
\eq
{\chi'''\,}^4_0  =  \tr ({\bq{2}{0}})^2 \qquad  (G=0)\ .
\eqn{chi''0-2}

Finally for $n=1$ one finds the solutions (there is none for $G=1$):
\eq
{\chi'\,}^3_0  = \tr c^3 \qquad  (G=0)\ ,
\eqn{chi0-1}
\eq
{\chi''\,}^3_0 = \tr (c^2\bq{1}{0})
     = -\dz \tr (c\bq{1}{0}) \qquad (G=0) \ ,
\eqn{chi'0-1}
and
\eq
{\chi'''\,}^3_0  =  \tr ({\bq{1}{0}})^3 \qquad  (G=0)\ .
\eqn{chi''0-1}

Climbing the ladder \equ{descent} from $\ch{n+2+G}{0}$
up to $\ch{G}{n+2}$ can be achieved through the repeated
application of the ``ladder climbing operator'' \cite{gms-plb2559165}
\eq
\na= dx^\m\hat\WW^{\rm S}_\m \ ,
\eqn{defnabla}
where $\hat\WW^{\rm S}_\m$ is the supersymmetry generator defined
by\footnote{The supersymmetry Ward operator $\wsh{\x}$
\equ{wardopsusyhat} defines a linear map, \ie a one-form,
from the vectors $\x$ to the space of the functional
differential operators. This is the intrinsic definition  of the
functional operator valued one-form $\na$.}
\eq
\wsh{\x} = \x^\m \hat\WW^{\rm S}_\m\ .
\eqn{w-mu-s}
The action of $\na$ on the ladder fields is given by:
\eq\ba{l}
\na\fah{A}{p} = (p+1)\fah{A}{p+1}\ ,\quad {\scr p= 0,\cdots,n+1}\ ,   \es
\na\fah{A}{n+2} = 0 \ ,
\ea\eqn{nabla}
It is a derivation, which commutes with the exterior derivative $d$.
The commutation rules of $\na$ with the other operators are:
\eq
[\dz,\na]=d\ ,\qquad
   [\wsh{\x},\na]=[\wte{\e},\na]=[\hat\GG,\na]=[\hat\FF,\na]=0\ ,
\eqn{alg-nabla}
from which it obviously follows that
\eq
[\d,\na]= d\ .
\eqn{delta,nabla}
These commutation rules are simple consequences of the algebra
\equ{alg-lin1}, which also implies the anticommutation rules
\eq
\lac \hat\WW^{\rm S}_\m,\hat\WW^{\rm S}_\n \rac =0\ ,
\eqn{anticommutation} from which follows the identity
\eq
\na^{n+2}\wsh{\x} = 0 \ .
\eqn{morepotency}

One easily checks now that, for a given zero-form $\ch{n+2+G}{0}$,
which is a solution of the last descent equation \equ{descent},
the forms
\eq
\ch{n+2+G-p}{p} = \frac{1}{p!}\na^p\, \ch{n+2+G}{0} \ ,\quad
       {\scr p=0,\cdots,n+2}\ ,
\eqn{laddersol}
do indeed solve the whole set of descent equations
\equ{descent}\footnote{See~\cite{gms-plb2559165}
for a previous version of this  construction.}.

We remark that, due to the first commutation rule
\equ{alg-nabla}, the solutions constructed from \equ{chi'0},
\equ{chi'0-2}  and \equ{chi'0-1} yield trivial cocycles, \ie the
corresponding insertions \equ{deltaq} are $\dz$-variations. On the other
hand the solutions derived from the zero-forms \equ{chi0},  \equ{chi0-2},
\equ{chi''0-2}, \equ{chi0-1} and \equ{chi''0-1} lead to non-trivial
$\dz$-cocycles.

Let us show that these are the only non-trivial solutions of \equ{cobz}.
It is clear that the problem of solving the descent equation
\eq
\dz\ch{g}{n+2+G-g} = d\chi^{g+1}_{n+1+G-g}\ ,
\qquad {\scr g=G,...,G+n+1}\ ,
\eqn{stepq}
for $\ch{g}{n+2+G-g}$ (a solution $\chi^{g+1}_{n+1+G-g}$
of the lower descent equation being given) is a problem
of local $\dz$-cohomology, \ie of solving the homogeneous equation
\eq
\dz \o =0 \ .
\eqn{localcohom}
The most convenient way to solve the latter equation
is to introduce a new filtration, with the counting operator
\eq
N=\sum_{{\rm all\ fields}\ \vf} \intr \tr\ \vf\fud{}{\vf}\ ,
\eqn{countingop}
as filtering operator. According to this filtration $\dz$ splits into
\eq
\dz = \d_{0,0} + \d_{0,1}\ , \quad{\rm with}\quad \d_{0,0}^2=0\ .
\eqn{dzz}
The action of the coboundary operator $\d_{0,0}$ on the
two field ladders $\fah{A}{}$, $A=1,2$ reads:
\eq\ba{l}
\d_{0,0} \fah{A}{0}=0\ ,   \es
\d_{0,0} \fah{A}{p} = d\fah{A}{p-1}\ ,\quad {\scr p=1,\cdots,n+2}\ .
\ea\eqn{dzzladder}
It is shown in~\cite{psor-nonren} that in the case of {\em complete}
field ladders -- as realized here -- the local cohomology of $\d_{0,0}$
depends only on the ladder components which are non-differentiated
zero-forms. These zero-forms in the present case are
$\fah{1}{0} =c$ and $\fah{2}{0}=\bq{n}{0}$. It follows that
the local $\d_{0,0}$-cohomology is empty in the space of forms of
degree $p>0$. The local $\dz$-cohomology, being isomorpic to a subset of
the local $\d_{0,0}$-cohomology, is empty as well in this space.
The consequence of this result is~\cite{lps}
that the general solution $\bar\chi^{g}_{n+2+G-g}$ of the
descent equations for a given non-trivial $\ch{n}{0}$ (see \equ{chi0})
is unique modulo--$\dz$ and modulo--$d$:
\eq
\bar\chi^{g}_{n+2+G-g} = \ch{g}{n+2+G-g} + \dz(\cdots) + d(\cdots)\ .
\eqn{sol-mod-mod}
Hence the whole $\dz$-cohomology for $G=0,1$ is given by all the linear
superpositions of the $\dz$-cocycles
\eq
\D^G_{(n+2)} = \intr \frac{1}{(n+2)!}\na^{n+2}
     \left. \prod_{k=1}^K
     \tr c^{n_k}\right|_{\sum_k n_k=n+2+G\ ,\quad n_k\ {\rm odd}}\ ,
     \quad{\scr n\ge 1}\ ,
\eqn{cohomology}
and, in addition for $n=2\ {\rm or}\ 1$, of the $\dz$-cocycles
\eq\ba{l}
\D^0_{(4)} = \dint_{\real^4}
      {1\over4!}\na^4\lp\tr(\bq{2}{0})^2\rp \ ,\es
\D^0_{(3)} = \dint_{\real^3}
      {1\over3!}\na^3\lp\tr(\bq{1}{0})^3\rp \ .
\ea\eqn{phi-terms}.

The cocycles \equ{cohomology} are solutions of the full cohomology
condition \equ{cocyceq}. This follows indeed from the
commutator \equ{delta,nabla} of the full coboundary operator
$\d$ with the climbing operator $\na$, from the identity
\equ{morepotency} and from the fact that $\prod\tr c^{n_k}$ is
annihilated by $\hat\GG$ \equ{ghosteq-hat} and $\hat\FF$
\equ{ghost-assoc-hat}, and is independent of $\e$ and $v$.
Moreover they cannot be $\d$-coboundaries since
they are independent of the global ghosts  $\e$, $\x$, $v$ and $u$.

The same argument applied to the cocycles \equ{phi-terms} shows
that these are not $\d$-invariant because of their dependence on
$B^n_0$, and thus must be rejected.

Hence \equ{cohomology} represents the whole $\d$-cohomology
for $G=0,1$ and $n\ge1$.
\subsection{Anomalies and invariant counterterms}
{\bf 1)} The solutions \equ{cohomology} with ghost number
$G=1$ yield all the possible anomalies of the $(n+2)$-dimensional
$BF$ systems. Some examples are shown in the next section.
The actual presence of anomalies depends on a case by case
investigation of the group theoretical structure of the model
and on explicit computations of the anomaly coefficients, possibly
with the help of a non-renormalisation theorem.

\noindent {\bf 2)}
The cocycles \equ{cohomology} with $G=0$ ghost number yield the
non-trivial counterterms, \ie those which generate the renormalization of
physical coupling constants. Trivial $\d$-invariant counterterms,
if any, of the form
\eq
\D_{\rm trivial}(\fah{1}{},\fah{2}{})
                = \d \D^-(\fah{1}{},\fah{2}{})\ ,
\eqn{trivial-ct}
where $\D^-$ is a local functional of ghost number $-1$,
would correspond to the renormalization of parameters
which can be compensated by field redefinitions.

Let us show that there is no such trivial counterterm.
In order for a trivial counterterm to be independent
of the global ghosts  $\x$, $\e$, $u$ and $v$,
$\D^-$ has to be independent from them too and it has to fulfill
the conditions
\eq
\wsh{\x}\D^- = 0\ ,
\eqn{ct-susy}
\eq
\hat\GG  \D^- =   \hat\FF  \D^- = 0\ .
\eqn{ct-ghosteq}
It is shown in Appendix~\ref{susyinvariants} that the general solution
of the supersymmetry condition \equ{ct-susy} reads
\eq
\D^- = \intr \na^{n+2}\O^{n+1}_0 \ .
\eqn{delta-}
The zero-form $\O^{n+1}_0$ is a superposition of the monomials
${\chi'\,}^{n+1}_0$ \equ{chi0} and of
\eq
{\chi^{''''}\,}^{n+1}_0  =  \tr (c\bq{n}{0}) \ .
\eqn{chi0b}
$\chi'$, yielding a $\D^-$ which is $\d$-invariant, leads to a
vanishing counterterm, whereas $\chi^{''''}$ leads to a $\D^-$
which does not obey \equ{ct-ghosteq} and thus must be discarded.

In conclusion the possible invariant counterterms are given
by the non-trivial cocycles  \equ{cohomology} with ghost number
$G=0$. Their actual occurrence has to be tested case by case as for the
anomalies.
\section{Discussion of the result: some examples}
The purpose of this section is to discuss in details the algebraic
structure of some $BF$ systems by using the general results on the
cohomology of the operator $\delta$. As explicit models we will analyze
the cases $d=4,5,6$ and $7$.
\subsection{Four-dimensional model}
Let us begin with the case $n=2$ which corresponds to a four dimensional
$BF$ system~\cite{gms-plb2559165}.

On one hand the non-trivial cocycle of ghost number one is given by
\eq\ba{rl}
\Delta^1_{(4)} & = {\displaystyle \frac{1}{5!}}
       {\displaystyle\int}_{\real^4} \na^4\, \tr c^5  \es
   &      = {\displaystyle\int}_{\real^4} \Lp  A^4c + {\hat b}^{-3}_4 c^4
 + {\hat b}^{-2}_3( Ac^3 + cAc^2 + c^2Ac + c^3A ) \es
  & {\ }{\ }{\ }{\ }
    + {\hat b}^{-1}_2 ( A^2c^2 + Ac^2A + c^2A^2
                + AcAc + cAcA + cA^2c ) \es
  & {\ }{\ }{\ }{\ }
    + {\hat b}^{-1}_2 {\hat b}^{-1}_2 c^3 +
      {\hat b}^{-1}_2 c {\hat b}^{-1}_2 c^2 \Rp
   \ .
\ea\eqn{cocy4expr}
and has the quantum numbers of a Slavnov anomaly. Actually it is
well known~\cite{zumino} that the ghost polynomial $\tr c^5$ is related
via descent equations to the non-abelian gauge anomaly
(or Adler-Bardeen-Bell-Jackiw anomaly)
\eq
\AA_{(4)} =\frac{1}{2}\int_{\real^4}\tr\Lp dc(AdA+dAA+A^3)\Rp\ .
\eqn{anom4}
The expression \equ{cocy4expr} is in fact nothing but the gauge anomaly
\equ{anom4} (modulo a $\BB$-coboundary) written in a way which is
compatible with the Ward identity of supersymmetry \equ{wardsusyshat}.
In this case however, the absence of anomalies is ensured by the fact
that all the fields belong to the adjoint representation of
the gauge group. Indeed this implies that the Feynman rules involve
only the structure constants $f^{[abc]}$. This forbids the appearence
of the totally symmetric tensor $d^{(abc)}$ which is present
in the expression \equ{cocy4expr}.

On the other hand there is no invariant counterterm. Thus the
four dimensional model is finite to all orders of perturbation theory.
\subsection{Five-dimensional model}
The general expression \equ{cohomology} shows that the ghost
polynomial $\tr c^5$ -- which gave the Slavnov anomaly in four
dimensions -- determines also the cohomology for the five-dimensional
$BF$ system (\ie $n=3$), but now in the sector of ghost number $G=0$.
The relevant cocycle for this case is
\eq
\Delta^0_{(5)}  = \frac{1}{5!} \int_{\real^5} \na^5\, \tr c^5  \ .
\eqn{bf5cocycle}
This cocycle, whose explicit form is given by
\eq\ba{rl}
\Delta^0_{(5)} = {\displaystyle\int}_{\real^5} & \tr \Lp
     \frac{1}{5}  A^5  + {\hat b}^{-4}_5 c^4
     + {\hat b}^{-3}_4 ( Ac^3 + c^2Ac + cAc^2 + c^3A ) \es
 &
     + {\hat b}^{-2}_3 ( {\hat b}^{-1}_2c^3 + c{\hat b}^{-1}_2c^2 +
                         c^2{\hat b}^{-1}_2c +c^3{\hat b}^{-1}_2 )  \es
 &
     + {\hat b}^{-1}_2 {\hat b}^{-1}_2 ( Ac^2 + cAc + c^2A )
     + {\hat b}^{-1}_2 ( A^3c + A^2cA + AcA^2 + cA^3 ) \es
 &
     + {\hat b}^{-1}_2A {\hat b}^{-1}_2c^2
     + {\hat b}^{-1}_2c {\hat b}^{-1}_2Ac
     + {\hat b}^{-1}_2c {\hat b}^{-1}_2cA  \es
 &
     +{\hat b}^{-2}_3 ( A^2c^2 + AcAc + Ac^2A + cA^2c
+ cAcA + c^2A^2 ) \Rp
   \ ,
\ea\eqn{cocy5expr}
has the quantum numbers of the five-dimensional $BF$ action
and corresponds
to a possible counterterm. Moreover one should note that this term, like
the cocycle \equ{cocy4expr}, contains the totally symmetric tensor
$d^{(abc)}$ which cannot be generated in a model containing
only fields in the adjoint representation. This means that the
expression \equ{cocy5expr} can appear as a counterterm only
if it has been included in the initial classical action. In this
case one has to deal with a more general model which contains
a generalized Chern-Simons term. It is not difficult in fact
to show that the above expression coincides, modulo
a $\BB$-coboundary, with the Chern-Simons five-form~\cite{zumino}
\eq
\AA_{(5)} =   \int_{\real^5} \tr \Lp
         AdAdA + \frac{2}{3}A^3dA + \frac{3}{5}A^5  \Rp  \ .
\eqn{chernsim5}
In this sense, this $BF$ system can be regarded as a higher dimensional
generalization of the three dimensional Chern-Simons theory. It is
apparent that the generalized Chern-Simons action can
be likewise  consistently included in any odd dimension.

But for the {\em pure} $BF$ theory there is no possible counterterm.
Since moreover no anomalies are allowed by the cohomology
(no cocycles with $G=1$), the pure $BF$ theory in five dimensions
is finite.
\subsection{Six- and seven-dimensional models}
Let us close this section by considering the six- and the
seven-dimensional (\ie $n=4, 5$) models. The relevant
ghost polynomials are given by $\tr c^7$ and $\tr c^3\tr c^5$
which allow to define the cocycles:
\eq
\Delta^1_{(6)}  = \frac{1}{7!} \int_{\real^6} \na^6\, \tr c^7  \ ,
\eqn{bf6cocycle}
in six dimensions, and
\eq
\D^0_{(7)} = \frac{1}{7!} \int_{\real^7} \na^7\, \tr c^7\ ,
\eqn{bf7cocycle}
\eq
\phantom{\D}\D^1_{(7)}  = \frac{1}{7!} \int_{\real^7}
\na^7\, \tr c^3\, \tr c^5  \ ,
\eqn{bf7cocycle'}
in seven dimensions. The expression \equ{bf6cocycle}  corresponds to
a possible Slavnov anomaly for the six-dimensional theory, while
expressions \equ{bf7cocycle} and \equ{bf7cocycle'} yield a generalized
Chern-Simons counterterm and a possible Slavnov anomaly in
the seven-dimensional case.

It is interesting to note that we cannot exclude the presence of the
six-dimensional anomaly \equ{bf6cocycle} or of the seven-dimensional
counterterm \equ{bf7cocycle} by using group theoretical arguments
as in the four- and five-dimensional cases. One easily sees, indeed,
that the expressions \equ{bf6cocycle}, \equ{bf7cocycle} contain
a totally symmetric tensor of rank four which could be generated by using
the structure constants. A detailed analysis of the Feynman
graphs which could contribute to these terms and
the possibility of an Adler-Bardeen non-renormalization
theorem is beyond the aim of this paper and will be reported
on in a future work.
\section{Conclusions}
We have shown the existence of a supersymmetric structure generated
by the BRS transformations and by a vector valued operator obeying,
together with the translations, a Wess-Zumino-like superalgebra.
In an appropriate linear gauge of the Landau type, the vector
operator yields a symmetry of the theory which is broken,
but only linearly.

We have given a complete classification of the possible BRS anomalies
and invariant counterterms. It has turned out that none
of them is present in the pure $BF$ models in space-time dimensions
four and five: these theories are finite. For arbitrary dimensions the
analysis of our general algebraic results cannot exclude the presence
of anomalies or of counterterms of the Chern-Simons type. Definite
conclusions would require explicit knowledge of the numerical
coefficients of the anomaly or of the counterterms, which one can
obtain by evaluating the contributing Feynman diagrams.

The classification of the anomalies and of the invariant counterterms
has been given by solving the BRS and supersymmetry cohomology,
constrained by the ghost equation. The explicit construction
of these objects has been performed by applying $d$ times ($d$ being
the dimension of space-time) an operator $\na$ to the zero-form
ghost cocycles \equ{chi0}. This operator $\na$, which is nothing else
than the supersymmetry generator, has a well-defined geometrical
meaning: mapping $p$-forms into $(p+1)$-forms, it naturally solves the
descent equations owing to the fact that its commutator with the
cohomology operator $\d$ yield the exterior derivative  as shown
by eq.\equ{delta,nabla}.

\noindent{\large{\bf Acknowledgments}}: We wish to thank M. Abud, J.P.
Ader, F. Brandt, L. Cappiello, F.  Delduc, S. Emery, F. Gieres,
N. Maggiore and R. Stora for useful discussions.
One of the authors (O.P.) would like to thank the Max-Planck-Institut
for Physics in Munich and its members for their kind hospitality
during part of this work.
\appendix
\newtheorem{defin}{Definition}[section]
\newtheorem{prop}{Proposition}[section]
\section{Appendix: Supersymmetry invariants}\label{susyinvariants}
We will show that the most general supersymmetric integrated
local functional of the ladder fields \equ{ladder-hat}
has the form \equ{delta-}. The result is stated in prop.~\ref{prop2}
below.

It is convenient to introduce a superspace formalism~\cite{axelrod}.
Superspace
is the set of points $z = (x^\m,\theta^\m)$, $x$ being a point of
$\real^{D}$ and $\{\theta^\m|\, {\scr \m=1,\cdots,D}\}$ a set of $D$
Grassmann coordinates. A superspace integration measure is provided by
\eq
\int dzf(z) = \int d^Dx d^D\theta f(z)
    = \int d^{D}x \frac{\pa}{\pa\theta^1} \cdots
          \frac{\pa}{\pa\theta^{D}} f(z)\ .
\eqn{s-integration}
\begin{defin}
Superfields are superspace local functionals  $\Psi(z)$ of the ladder
fields $\fah{A}{p}$, transforming under supersymmetry
\equ{wardopsusyhat} as
\eq
\wsh{\x}\Psi(z) = \x^\m \frac{\pa}{\pa\theta^\m}\Psi(z)
\eqn{superfield}
\end{defin}
Superfields build up an algebra, a basis of which is provided by the
superfields
\eq
\F^{(A)}(z) = \sum_{p=0}^{D}\frac{1}{p!} \theta^{\m_1}\cdots\theta^{\m_p}
   \f^{(A)}_{\m_1\cdots\m_p}(x) \ ,
\eqn{superladders}
and their derivatives, where $\f^{(A)}_{\m_1\cdots\m_p}(x)$
is a component of the $p$-form $\fah{A}{p}$. According to the
transformation rules \equ{wardopsusyhat} $\F^{(A)}$ as well as its
$x$- and $\theta$-derivatives indeed tranform as superfields.
\begin{prop}\label{prop1} Every local functional $\D$
of the ladder fields, invariant under supersymmetry, \ie obeying
\eq
\wsh{\x}\D=0\ ,
\eqn{s-invar}
may be expressed as the superspace integral
\eq
\D=\int dz \Psi(z)
\eqn{s-integr}
of a composite superfield $\Psi$ made out of the superfields
$\F^{(A)}$ and their derivatives.
\end{prop}
\noindent{\bf Proof}: The most general $\D$ (of degree $N$ in the
fields) may be written as a multiple integral (\cf~\cite{pss}):
\eq
\D = \int dz_1\cdots dz_N\, \F^{(A_1)}(z_1)\cdots\F^{(A_N)}(z_N)
    K_{A_1\cdots A_N}(z_1,\cdots,z_N)\ .
\eqn{multint}
Due to the locality hypothesis and to translation invariance, the kernel
$K$ is a linear combination of products of derivatives of Dirac
distributions $\d^D(x_i-x_1)$ with $\theta$-dependent coefficients.
The invariance condition \equ{s-invar} is equivalent to the condition
\eq
\lp\sum_{i=1}^N \pad{}{\theta_i^\m}\rp K = 0
\eqn{cond-kernel}
for the kernel. The latter condition implies that the kernel
depends on the $\theta$'s only through their differences
$\theta_i-\theta_1$. Then it is easy to see that its most general
expression consists in a sum of terms of the form
\eq
 K_{A_1\cdots A_N}(z_1,\cdots,z_N) = \prod_{a=2}^N
  \pad{}{\theta_a^{\m_1}}\cdots \pad{}{\theta_a^{\m_{M_a}}}
    \pad{}{x_{a\m_1}}\cdots \pad{}{x_{a\m_{M_a}}}
        \Box^{N_a}\d(z_a-z_1)\ ,
\eqn{kernel}
where we have used the superspace delta distribution:
\eq\ba{l}
\dint dz_2\d(z_1-z_2)f(z_2) = f(z_1)\ ,\es
\d(z_1-z_2)= \lp \prod_{\m=1}^D(\theta_1^\m - \theta_2^\m)\rp
          \d^D(x_1-x_2)\ .
\ea\eqn{superdelta}
Introducing this result in \equ{multint} we conclude that $\D$
is a sum of terms of the form \equ{s-integr}
with
\eq
\quad \Psi(z) = \F^{(A_1)}(z)\prod_{a=2}^N
  \pad{}{\theta^{\m_1}}\cdots \pad{}{\theta^{\m_{M_a}}}
    \pad{}{x_{\m_1}}\cdots \pad{}{x_{\m_{M_a}}}
        \Box^{N_a} \F^{A_a}(z)
\eqn{s-integrant}
is a superfield.
\begin{prop}\label{prop2}
The supersymmetric local functional of Prop.~\ref{prop1} can be written
as the space-time integral
\eq
\D = \int_{\real^D} \na^D\O_0\ ,
\eqn{nabla-d}
where $\O_0$ is a zero-form and $\na$ the ladder climbing operator
defined in eqs \equ{defnabla}, \equ{w-mu-s}.
\end{prop}
\noindent{\bf Proof}: This follows from the result \equ{s-integr}, the
definition \equ{s-integration} of superspace integration and from
the supersymmetry transformation law \equ{superfield}. The zero-form
$\O_0$ is proportional to the $\theta=0$ component of the superfield
$\Psi(z)$.

\end{document}